\documentclass{iau} 
\usepackage{graphicx, hyperref}

\newcommand{\msol}{$M_{\odot}$}

\title[Variable stars in the Gaia era] 
{Variable stars in the Gaia era: \\ Mira, RR Lyrae, $\delta$ and Type-II Cepheids}

\author[Martin A.T. Groenewegen]   
{Martin A.T. Groenewegen}

\affiliation{Koninklijke Sterrenwacht van Belgi\"e \\ 
Ringlaan 3, B-1180 Brussel, Belgium \\ email: {\tt martin.groenewegen@oma.be} }

\pubyear{2017}
\volume{330}  
\jname{Astronomy and Astrophysics in the Gaia sky}
\editors{A. Recio-Blanco, P. de Laverny, A. Brown \& T. Prusti, eds.}
\begin{document}

\maketitle

\begin{abstract}
Classical variables like RR Lyrae, classical and Type-II Cepheids and Mira variables all follow
period-luminosity relations that make them interesting as distance indicators. 
Especially the RR Lyrae and $\delta$ Cepheids are crucial in establishing the distance scale 
in the Universe, and all classes of variables can be used as tracers of galactic structure.
I will present an overview of recent period-luminosity relations and review the work that has 
been done using the {\it Gaia} DR1 data so far, and discuss possibilities for the future.
\end{abstract}

\firstsection 
\section{Introduction}

Miras, RR Lyrae, classical and Type-II Cepheids belong to the oldest known variable stars, certainly
in the literature of the west.
The American Association of Variable Star Observers (AAVSO) website has interesting historical 
information about the prototypes
$o$ Ceti or Mira (discovered by Fabricius in 1596, see \url{www.aavso.org/vsots\_mira}),
RR Lyrae (discovered by Wilhelmina Fleming, and published in Pickering et al. 1901, see \url{www.aavso.org/vsots\_rrlyr}), 
and $\delta$ Cephei (discovered by John Goodricke in 1784, see \url{www.aavso.org/vsots\_delcep}).

Type-II Cepheids are subdivided in three classes, typically based on pulsation period: 
The BL Herculis variables (BLH; periods 1-4 days; discovery paper by Cuno Hoffmeister 1929),
the W Virginis stars (WVir; periods 4-20 days; discovered by Eduard Sch\"onfeld in 1866, see \url{www.aavso.org/vsots\_wvir}),
and the RV Tauri stars (RVT; periods 20-$\sim$70 days; discovered by Lidiya Tseraskaya (or Ceraski), 
published in Ceraski 1905, see \url{www.aavso.org/vsots\_rvtau}).

In evolutionary terms, RR Lyrae (RRL) variables are evolved, metal poor, core He-burning stars at or
slightly brighter than the zero-age horizontal branch (ZAHB). Marconi et al. (2015) provide recent
nonlinear, time-dependent convective hydrodynamical models of RRL over a broad range in metal 
abundances ($Z$ = 0.0001-0.02) and masses, ranging from 0.8~\msol\ (for $Z$= 0.0001) to 0.54~\msol\ (for $Z$= 0.02).
They provide analytical relations for the edges of the instability strip (IS) as a function of $Z$.
Period-radius-metallicity relations for fundamental and first-overtone pulsators are determined, as well as
a large set of period-luminosity and period-Wesenheit relations.

Classical or $\delta$ Cepheids (CEPs) are evolved objects with initial masses in the range $\sim2$ to $\sim15$~\msol.
Theoretical pulsation models have been calculated by Bono et al. (2000) and Anderson et al. (2014, 2016) who
considered the effect of rotation on the evolution and pulsation.
A Cepheid can cross the IS up to three times
The, so-called, first crossing occurs when the star evolves from the main sequence to the red giant branch 
during a core contraction phase.
This crossing is expected to be fast, and Cepheids in this phase should  be rare.
The majority of Cepheids are expected to be on the second and third crossings during the 
so-called "blue loops" experiencing core helium burning.
%


As mentioned above, the Type-II Cepheids (T2C) are subdivided in three classes, typically based on period,
but they are thought to have different evolutionary origins. 
Evolutionary modelling of T2Cs has been pioneered by Gingold (1976, 1985) establishing the classical picture that 
T2Cs are low-mass stars, evolving from the blue HB through the IS to the asymptotic giant branch (AGB) for the 
short-period stars, blue loops off the AGB for the stars of intermediate period, and post-AGB (PAGB) evolution for 
the longest period, also see Wallerstein (2002) and Bono et al. (2016).

The anomalous Cepheids (ACs) are also pulsating stars which overlap in period range with RRL and the BLH stars. 
They form a separate $PL$ relation clearly different from the RRL, classical Cepheids and T2Cs.
They pulsate in the fundamental mode (FU) and first overtone (FO) mode (unlike T2C).
Models have been calculated by Fiorentino \& Monelli (2012). Their mean mass is around $1.2 \pm 0.2$~\msol, and there
is also discussion if ACs are the result of binary interaction.

Groenewegen \& Jurkovic (2017a, b) recently studied the 335 T2C and ACs discovered by the OGLE-III survey in the Large and 
Small Magellanic Cloud. From fitting the spectral energy distribution (SED) they derived effective temperature and luminosity.
In the 2017a paper the resulting Hertzsprung-Russell diagram was compared in a qualitative way to modern evolutionary tracks. 
In agreement with the findings cited above the BL Her can be explained by stars in the mass range 
$\sim 0.5-0.6$~\msol\ and the ACs by stars in the mass range $\sim 1.1-2.3$~\msol.
The origin of the (p)WVir is unclear however: tracks of $\sim 2.5-4$~\msol\ cross the IS at the correct luminosity, as well as
(some) lower mass stars on the AGB that undergo a thermal pulse when the envelope mass is small, but the timescales make these
unlikely scenarios to explain this class of objects as a whole. The peculiar W Vir have been suggested to be binaries, and
in general, some of the phenomenon observed in T2C and ACs may well be linked to so called binary evolutionary 
pulsators (BEP; Karczmarek et al. 2016).

In the 2017b paper, pulsation models for RRL (Marconi et al. 2015) and Cepheids (Bono et al. 2000) were used to estimate the 
pulsation mass for all objects. Both estimates agreed best for the BLH ($M \sim 0.49$ \msol) and the ACs ($M \sim 1.3$ \msol).
The masses of the W Vir appeared similar to the BL Her. The situation for the pWVir and RVT stars was less clear. 
For many RV Tau the masses are in conflict with the standard picture of (single-star) post-AGB evolution, the
masses being either too large ($\gtrsim$ 1 \msol) or too small ($\lesssim$ 0.4 \msol).

Groenewegen \& Jurkovic (2017a) found that $\sim60\%$ of the RVT showed an infrared excess in their SEDs, 
not unsurprising if RVT have indeed evolved of the AGB. 
Surprisingly however, $\sim 10\%$ of the W Vir (including the pWVir) objects also showed an infrared excess,
confirming the result of Kamath et al. (2016) that there exist stars with luminosities below that predicted 
from single-star evolution that show a clear infrared excess, and which they called dusty post-red giant branch
stars, and suggested to have evolved off the RGB as a result of binary interaction.

AGB and super-AGB stars are intermediate mass stars (initial mass $\sim$0.8-12~\msol) in the last phase of
active nuclear burning, that undergo double-shell burning, experience thermal pulses (or Helium shell flashes) that 
change the composition of the envelope, making it increasingly rich in carbon, so that S-stars 
(C/O ratio close to one, and that show increased abundances of $s$-process elements), and C-stars (Carbon stars with C/O$>$1)
can form. They are cool giants, where dust can form close to the star that is driven outward in a slow stellar wind.

(S)AGB stars also pulsate, classically divided into irregular (Lb), semi-regular (SR) and Mira (M) variables.
The SR and M are sometimes also called long-period variables (LPVs), as they are not so different.
Miras are not necessarily less regular than the SR, and the definition that Miras should have an pulsation
amplitude in the visual band larger than 2.5 magnitudes is arbitrary.

That Miras follow a $PL$ relation is well known (Glass \& Lloyd Evans 1981), and it should be noted 
that a 500 day Mira is $\sim$0.5 mag brighter than a 50 day Cepheid in the near and mid-IR.
The revolution came with advent of the microlensing surveys, MACHO and OGLE.
Wood et al. (1999) and Wood (2000) showed that red giants in the LMC follow several sequences, 3 that define
pulsating stars, a sequence that consists of binary systems, and one that is formed by the long secondary period (LSP)
that occurs in many red giants.
Subsequent works expanded on this in various ways (Ita et al. 2004, 
Soszy\'nski et al. 2004, 2005, 
Fraser et al. 2008, Riebel et al. 2012, Soszy\'nski \& Wood 2013, Soszy\'nski et al. 2013) and revealed many 
more (sub-)sequences, including those for RGB stars.

\vspace{-2.0mm}
\section{Cosmological connection}

The small dispersion in the $PL$ relation of classical Cepheids makes them the primary calibrator in the distance 
ladder, and ultimately in determining the Hubble constant (Freedman et al. 2001). Riess et al. (2016) find 
$H_0= 73.24 \pm 1.74$ km/s/Mpc by using locally calibrated Cepheids (15 Cepheids with parallaxes in our MW, 
8 detached eclipsing binaries (dEBs) in the LMC, 2 dEBs in M31, and the megamaser in NGC 4258), to determine 
the brightness of Type-I SNe in 19 galaxies that host Cepheids and SNIa, and then measure the distance to 
$\sim300$ SNIa in the Hubble flow with  $z<0.15$.
This value for $H_0$ differs by more than 3$\sigma$ from the $H_0= 66.93 \pm 0.62$  km/s/Mpc determined 
by the Planck mission (Planck Collaboration et al. 2016).
Whether this discrepancy is real is of obvious importance and requires that all steps in the stellar distance 
ladder are investigated and improved, and this includes the Cepheid $PL$ relation.

In this line, the Carnegie-Chicago Hubble Program (Beaton et al. 2016; this volume) aims at a 3\%  measurement 
of $H_0$ using alternative methods to the traditional Cepheid distance scale. They aim to establish a 
completely independent route to the Hubble constant using RRL variables, and the tip of the red giant branch (TRGB) method.
This requires a reassessment of the RRL $PL$ relation.

\vspace{-2.0mm}
\section{Period-luminosity relations}

Mostly recent empirical $PL$ and $PLZ$ relations for RRL, Type-II, ACs and classical Cepheids in 
selected filters and Wesenheit relations are compiled in Table~\ref{Tab-PL}. 
If a slope was adopted it is listed between parentheses.
Large the table is, it is certainly not complete and the numbers hide important details in their derivation.
Period-luminosity relations exist in other infrared filters than $K$, and in other Wesenheit combinations
than $V$ and $I$, or $V$ and $K$.
The original references should be consulted about solutions for other filter combinations or pulsation modes, 
the details of the filter(s) used, the details in the definition of the Wesenheit function, 
any cuts in pulsation period that were applied, or, mostly for the LMC, whether the effect 
of the orientation of the disc was taken out or not.

Table~\ref{Tab-PL} includes solutions based on {\it Gaia} (Gaia Collaboration et al. 2016b) data release one 
(GDR1; Gaia Collaboration et al. 2016a).
Clementini et al.\ (2016) derived $PL$ relations in the {\it Gaia} $G$ band based on data in the south ecliptic pole
in the outskirts of the LMC.
Gaia collaboration et al.\ (2017) contains several $PL$ relations based on known RRL, T2C and CEP in our Galaxy 
based on the TGAS solution.
They present solutions based on three approaches.
The first is based on a least-square fit of absolute magnitude versus $\log P$ where the absolute magnitude is calculated from
\begin{equation}
M = m_0 + 5 \log \pi -10,
\label{Eq-DM}
\end{equation}
with $m_o$ the dereddened magnitude and the parallax is in milli-as.
A simple application of this method requires a selection in parallax space ($\pi > 0$) and is therefore subject to
Lutz-Kelker bias (Lutz \& Kelker 1973, Oudmaijer et al. 1998, Koen 1992), which 
Gaia collaboration et al.\ (2017) did not correct for. 
They also present two methods that work in parallax space.
In this case Eq.~\ref{Eq-DM} is rewritten (for a $PL$ relation $\alpha + \beta \log P$, or similarly for 
a $M_{\rm V}-$[Fe/H] relation for RRL) as
\begin{equation}
10^{0.2 \alpha} = \pi \cdot 10^{0.2 (m_0 -\beta \log P -10)}.
\end{equation}
The first method is based on a Bayesian approach, and the other, on a weighted non-linear least squares solution of this
equation, called the astrometric based luminosity (ABL), and these are the 
solutions listed in Table~\ref{Tab-PL}. They cite Arenou \& Luri (1999), although the method was used in a classical paper
by Feast \& Catchpole (1997) to determine the zeropoint of the Cepheid $PL$ relation based on {\it Hipparcos} data
for 220 Cepheids. The method was shown to be free from bias by Koen \& Laney (1998).

Table~\ref{Tab-PL} also includes $PL$ relations based on individual distances to Galactic and MC Cepheids
based on the Baade-Wesselink method (Storm et al. 2011a,b, Groenewegen 2013).
This method depends on the so-called projection factor, $p$, that translates the pulsational velocity to
the radial velocity in the line-of-sight measured via spectroscopy.
Both papers derive a $p$ factor that depends quite strongly on period from the condition that the distance to the LMC
should not depend on pulsation period (Storm et al. find $p = 1.550 -0.186 \log P$; Groenewegen 2013 find
$p= 1.50 -0.24 \log P$). However, the most recent studies  indicate that
the data is consistent with a constant $p$ factor of 1.29 $\pm$ 0.04 (Kervella, this volume; Kervella et al. 2017).
The reason behind this discrepancy is currently unknown.


\begin{table}
  \begin{center}
  \caption{PLZ relations for RRL and different classes of Cepheids \newline \hspace{\textwidth} 
           (mag =  $\alpha + \beta \log P + \gamma$ [Fe/H]). }

  \label{Tab-PL}
 {\scriptsize
  \begin{tabular}{|l|c|c|c|c|c|c|}\hline 

Class &    Band    &      $\alpha$      &      $\beta$     &       $\gamma$     &  Sample & Reference  \\
\hline
RRLab & V          & $19.385 \pm 0.017$ &         -        & $0.214 \pm 0.047$ & LMC & Gratton et al. (2004) \\ 
RRLab & $M_{\rm V}$ &   $0.93 \pm 0.12$  &         -        &  $0.23 \pm 0.04$  & GCC & Chaboyer et al. (1999) \\ %
RRLab & $M_{\rm V}$ &   $0.82 \pm 0.04$  &         -        & ($0.214$)         & GAL & Gaia Collaboration et al. (2017) \\ 
RRLab & W(V,I)     & $17.172 \pm 0.003$ & $-2.933 \pm 0.009$ &      -        & LMC & Jacyszyn-Dobrzeniecka et al. (2017) \\
RRLab & W(V,I)     & $17.492 \pm 0.007$ & $-3.001 \pm 0.028$ &      -        & SMC & Jacyszyn-Dobrzeniecka et al. (2017) \\
RRLab & K          &  $17.43 \pm 0.01$  & $-2.73 \pm 0.25$ &  $0.03 \pm 0.07$  & LMC & Murareva et al. (2015) \\ 
RRLab & K          &  $13.28 \pm 0.02$  & $-2.33 \pm 0.08$ &         -         & M5  & Coppola et al. (2011) \\ 
RRLab & K          & $10.420 \pm 0.024$ & $-2.33 \pm 0.07$ &        -          & M4  & Braga et al. (2015) \\ 
RRLab & K          & $12.752 \pm 0.054$ & $-2.232 \pm 0.044$ & $0.141 \pm 0.020$  &  $\omega$ Cen  & Navarrete et al. (2017) \\ 
RRLab & $M_{\rm K}$ &  $-1.16 \pm 0.27$  & ($-2.33$)        &         -         & GAL & Groenewegen \& Salaris (1999) \\ 
RRLab & $M_{\rm K}$ &  $-1.05 \pm 0.13$  & $-2.38 \pm 0.04$ &  $0.08 \pm 0.11$  & GCC & Sollima et al. (2006) \\ %
RRLab & $M_{\rm K}$ &  $-0.95 \pm 0.14$  & $-2.53 \pm 0.36$ &  $0.07 \pm 0.04$  & GAL & Murareva et al. (2015) \\ 
RRLab & $M_{\rm K}$ &  $-1.17 \pm 0.10$  & ($-2.73$)        &  $0.07 \pm 0.07$  & GAL & Gaia Collaboration et al. (2017) \\ 

RRLab & [3.6]      & $10.229 \pm 0.010$ & $-2.332 \pm 0.106$ &       -         & M4   & Neeley et al. (2015) \\ 
RRLab & [4.5]      & $10.192 \pm 0.010$ & $-2.336 \pm 0.105$ &       -         & M4   & Neeley et al. (2015) \\ 

RRLab & W1         & $-1.113 \pm 0.013$ &  $-2.38 \pm 0.20$  &       -         & GAL  & Klein et al. (2014) \\ 
RRLab & W2         & $-1.111 \pm 0.013$ &  $-2.39 \pm 0.20$  &       -         & GAL  & Klein et al. (2014) \\ 
&&&&&& \\

T2C   & G           & $18.640 \pm 0.085$ & $-1.650 \pm 0.109$ &       -         & LMC  & Clementini et al. (2016) \\ 
T2C   & W(V,I)      & $17.365 \pm 0.015$ & $-2.521 \pm 0.022$ &       -         & LMC  & Matsunaga et al.  (2009) \\ 
T2C   & W(V,I)      & $17.554 \pm 0.083$ & $-2.304 \pm 0.107$ &       -         & SMC  & Matsunaga et al.  (2011) \\ 
T2C   & K           &  $13.27 \pm 0.10$  &  $-2.24 \pm 0.14$  &       -         & GB   & Groenewegen et al.  (2008) \\ 
T2C   & K           & $17.412 \pm 0.029$ & $-2.278 \pm 0.047$ &       -         & LMC  & Matsunaga et al. (2009) \\ 
T2C   & K           & $17.600 \pm 0.082$ & $-2.113 \pm 0.105$ &       -         & SMC  & Matsunaga et al. (2011) \\ 
T2C   & K           &  $17.47 \pm 0.02$  & $-2.385 \pm 0.030$ &       -         & LMC  & Ripepi et al. (2015) \\ 
T2C   & K           & $17.405 \pm 0.038$ & $-2.483 \pm 0.089$ &       -         & LMC  & Bhardwaj et al. (2017) \\ 
T2C   & $M_{\rm K}$  &  $-1.58 \pm 0.17$  & ($-2.385$)         &       -         & GAL  & Gaia Collaboration et al. (2017) \\ 
T2C   & W(V,K)      &  $17.33 \pm 0.02$  &  $-2.49 \pm 0.03$  &       -         & LMC  & Ripepi et al. (2015) \\ 
T2C   & W(V,K)      & $17.415 \pm 0.012$ & $-2.456 \pm 0.025$ &       -         & LMC  & Bhardwaj  et al. (2017) \\ 
&&&&&& \\

AC FU & G         &  $18.00 \pm 0.04$  &  $-2.95 \pm 0.27$  &       -         & LMC  & Clementini et al. (2016) \\ 
AC FU & K         &  $16.74 \pm 0.02$  &  $-3.54 \pm 0.15$  &       -         & LMC  & Ripepi et al. (2014) \\ 
AC FU & W(V,K)    &  $16.58 \pm 0.02$  &  $-3.58 \pm 0.15$  &       -         & LMC  & Ripepi et al. (2014) \\ 
&&&&&& \\

CEP FU & G          & $17.361 \pm 0.020$ & $-2.818 \pm 0.032$ &       -         & LMC  & Clementini et al. (2016) \\ 
CEP FU & $M_{\rm V}$ &  $-1.43 \pm 0.10$  & ($-2.81$)          &       -         & GAL  & Feast \& Catchpole (1997) \\ 
CEP FU & $M_{\rm V}$ & $-1.275 \pm 0.023$ & $-2.678 \pm 0.076$ &       -         & GAL  & Fouqu\'e et al (2007) \\ 
CEP FU & $M_{\rm V}$ &  $-1.54 \pm 0.10$  &  ($-2.678$)        &       -         & GAL  & Gaia Collaboration et al. (2017) \\ 
CEP FU & W(V,I)     & $16.375 \pm 0.014$ & $-3.314 \pm 0.020$ &       -         & SMC  & Ngeow et al. (2015a)  \\ 
CEP FU & W(V,I)     & $15.897 \pm 0.001$ & $-3.327 \pm 0.001$ &       -         & LMC  & Inno et al. (2016)  \\ 
CEP FU & W(V,I)     & $16.492 \pm 0.002$ & $-3.358 \pm 0.005$ &       -     & SMC  & Jacyszyn-Dobrzeniecka et al. (2017) \\
CEP FU & W(V,I)     & $15.888 \pm 0.004$ & $-3.313 \pm 0.006$ &       -     & LMC  & Jacyszyn-Dobrzeniecka et al. (2017) \\
CEP FU & $M_{\rm W(V,I)}$ & $-2.60 \pm 0.03$ & $-3.32 \pm 0.08$ &      (0.0)      & MC+G  & Storm et al. (2011b) \\
CEP FU & $M_{\rm W(V,I)}$ & $-2.414 \pm 0.022$  & $-3.477 \pm 0.074$ &  -         & GAL  & Fouqu\'e et al. (2017) \\ 
CEP FU & $M_{\rm W(V,I)}$ &  $-2.82 \pm 0.11$  & ($-3.477$)     &       -         & GAL  & Gaia Collaboration et al. (2017) \\ 
CEP FU & K          & $16.494 \pm 0.026$ & $-3.212 \pm 0.033$ &       -         & SMC  & Groenewegen  (2000) \\ 
CEP FU & K          & $16.514 \pm 0.025$ & $-3.213 \pm 0.032$ &       -         & SMC  & Ngeow et al. (2015a) \\ 
CEP FO & K          & $15.941 \pm 0.032$ & $-3.132 \pm 0.083$ &       -         & SMC  & Bhardwaj et al. (2016b) \\ 
CEP FU & K          & $16.051 \pm 0.050$ & $-3.281 \pm 0.040$ &       -         & LMC  & Persson et al. (2004) \\ 
CEP FU & K          & $16.070 \pm 0.017$ & $-3.295 \pm 0.018$ &       -         & LMC  & Ripepi et al. (2012) \\ 
CEP FU & K          & $15.984 \pm 0.017$ & $-3.228 \pm 0.004$ &       -         & LMC  & Macri et al. (2015) \\ 
CEP FO & K          & $15.458 \pm 0.014$ & $-3.257 \pm 0.023$ &       -         & LMC  & Macri et al. (2015) \\ 
CEP FU & $M_{\rm K}$ & $-2.282 \pm 0.019$ & $-3.365 \pm 0.063$ &       -         & GAL  & Fouqu\'e et al. (2007) \\ 
CEP FU & $M_{\rm K}$ &  $-2.63 \pm 0.10$  & ($-3.365$)         &       -         & GAL  & Gaia Collaboration et al. (2017) \\ 
CEP FU & $M_{\rm K}$ &  $-2.33 \pm 0.03$  &  $-3.30 \pm 0.06$  &      (0.0)      & MC+G  & Storm et al. (2011b) \\ 
CEP FU & $M_{\rm K}$ &  $-2.49 \pm 0.08$  &  $-3.07 \pm 0.07$  & $-0.05 \pm 0.10$ & MC+G  & Groenewegen  (2013) \\ 
CEP FU & W(V,K)     & $15.870 \pm 0.013$ & $-3.325 \pm 0.014$ &       -         & LMC  & Ripepi et al. (2012) \\ 
CEP FU & W(V,K)     & $15.894 \pm 0.002$ & $-3.314 \pm 0.002$ &       -         & LMC  & Inno et al. (2016) \\ 
CEP FU & W(V,K)     & $15.837 \pm 0.049$ & $-3.287 \pm 0.010$ &       -         & LMC  & Bhardwaj et al. (2016a) \\ 
CEP FU & $M_{\rm W(V,K)}$ & $-2.87 \pm 0.10$ & ($-3.32$)        &       -          & GAL  & Gaia Collaboration et al. (2017) \\ 
CEP FU & $M_{\rm W(V,K}$ & $-2.69 \pm 0.08$ & $-3.11 \pm 0.07$ & $+0.04 \pm 0.10$ & MC+G  & Groenewegen  (2013) \\ 

CEP FU & [3.6]      &  $16.01 \pm 0.02$  &  $-3.31 \pm 0.05$ &        -         & LMC   & Monson et al. (2012) \\ 
CEP FU & [4.5]      &  $15.90 \pm 0.02$  &  $-3.21 \pm 0.06$ &        -         & LMC   & Monson et al. (2012) \\ 

CEP FU & $M_{\rm [24]}$ & $-2.46 \pm 0.10$ &  $-3.18 \pm 0.10$ &        -         & GAL   & Ngeow et al. (2015b) \\ 

\hline
  \end{tabular}
  }
 \end{center}
\end{table}

\vspace{-2.0mm}
\section{GCVS $\Longleftrightarrow$ GDR1}

As a usefull exercise I cross correlated the latest edition of the General Catalog of Variable Stars 
(version 5.1; Samus et al. 2017, see \url{www.sai.msu.su/gcvs/gcvs/gcvs5/htm/}) with the GDR1 for several types of variables.
The results are summarised in Table~\ref{Tab-GCVS}. 
The third column lists which identifiers were used in the search, 
the fourth column how many of those types are listed in the GCVS,
the fifth column how many have a parallax listed in the GDR1 TGAS solution, 
and the last column how many of those have a relative parallax error less than 16\%.
The week after the conference, on May 1st, Gaia Collaboration et al. (2017) appeared that did a very similar search, 
to ultimately derive the $PL$ relations listed in Table~\ref{Tab-PL}.

\begin{table}
  \begin{center}
  \caption{Link between GCVS classes and  GDR1. }

\label{Tab-GCVS}

 {\scriptsize
  \begin{tabular}{|l|c|c|c|c|c|c|}\hline 

Class  &  Type   &  GCVS        & Number  & Number  &       Number   \\
       &         &              & in GCVS & in GDR1 &         with  \\
       &         &              &         &         &  $(\sigma_{\pi}/\pi) < 0.16$ \\
\hline
RRL   &          & RRab, RRc                             &  6631 & 331 & 18 \\
T2C   & BLH/WVir & CW, CW:, CWA, CWA:, CWB, CWB:         &   271 &  44 &  2 \\ 
T2C   & RVT      & RV, RV:, RVA, RVA:, RVB, RVB+EA, RVB: &   159 &  52 &  0 \\
AC    &          & BLBOO                                 &     1 &   0 &  0 \\
CEP   &          & DCEP                                  &   632 & 289 &  1 \\
M/SR  &          & M, SRA, SRB                           & 10491 & 732 &  1   \\

\hline
  \end{tabular}
  }
 \end{center}
\end{table}

The RRL is the class with the largest number of (accurate) parallax data available.
As they are intrinsically faint GDR1 is not hampered by the fact that many bright stars are not listed there.
Table~\ref{Tab-RRL} lists the 18 RRL with $\sigma_{\pi} / \pi < 0.16$ plus SU Dra, sorted by relative parallax error.
SU Dra is included as it is one of five RRLs which have the parallax determined using the {\it HST} by Benedict et al. (2011).
For those five stars their is good agreement between the {\it HST} based parallaxes and GDR1.
The last column gives the parallax listed by van Leeuwen (2007) based on the re-reduction of the {\it Hipparcos} data.

The next entries in Tables~\ref{Tab-GCVS} and \ref{Tab-RRL} are for the T2C, of the BLH \& WVir, and RVT types.
Only 2 have an accurate parallax determined in GDR1.
$\kappa$ Pav, one of two T2C with an {\rm HST} based parallax in Benedict et al. (2011), is missing, probably because
it is so bright.
Interestingly, the parallax measurement for VY Pyx differs quite a bit from the {\rm HST} and the {\it Hipparcos} based value.

The ACs are listed under the identifier "BLBOO" in the GCVS.  There is only one, BL Boo, which
is not listed in GDR1.

 \begin{table}
  \begin{center}
  \caption{Data on RR Lyrae and Type-II Cepheids.}
  \label{Tab-RRL}
 {\scriptsize
  \begin{tabular}{|r|r|r|c|r|c|c|c|}\hline 

Name      & Hipparcos & Type &  Period &  $G$   & $\pi \pm \sigma_{\pi}$ & $\pi \pm \sigma_{\pi}$ & $\pi \pm \sigma_{\pi}$  \\
          &           &      &   (d)   &  (mag) &  (mas, GDR1 )         &     (mas, HST)        &   (mas, Hipp)          \\ \hline

\multicolumn{8}{c}{RR Lyrae} \\
RR Lyr    &  95497 & RRab & 0.567 &  7.6 & 3.64 $\pm$ 0.23 & 3.77 $\pm$ 0.13 & 3.46 $\pm$ 0.64 \\
FO CVn    &        & RRc  & 0.284 & 10.8 & 3.15 $\pm$ 0.25 &                 &                 \\
RZ Cep    & 111839 & RRc  & 0.309 &  9.2 & 2.65 $\pm$ 0.24 & 2.54 $\pm$ 0.19 & 0.59 $\pm$ 1.48 \\
CS Eri    &  12199 & RRc  & 0.311 &  8.9 & 2.16 $\pm$ 0.23 &                 & 2.71 $\pm$ 1.10 \\
X Ari     &  14601 & RRab & 0.651 &  9.5 & 2.02 $\pm$ 0.22 &                 & 0.88 $\pm$ 1.32 \\
UV Oct    &  80990 & RRab & 0.542 &  9.5 & 2.02 $\pm$ 0.23 & 1.71 $\pm$ 0.10 & 2.44 $\pm$ 0.81 \\
AR Per    &  19993 & RRab & 0.425 & 10.3 & 1.99 $\pm$ 0.24 &                 & 0.93 $\pm$ 1.45 \\
DX Del    & 102593 & RRab & 0.472 &  9.9 & 1.66 $\pm$ 0.22 &                 & 0.77 $\pm$ 1.38 \\
EW Cam    &  36213 & RRab & 0.628 &  9.4 & 1.69 $\pm$ 0.23 &                 & 2.13 $\pm$ 1.10	\\

V1057 Cas &        & RRc  & 0.423 & 10.0 & 2.20 $\pm$ 0.31 &                 &                 \\
XZ Dra    &  94134 & RRab & 0.476 & 10.3 & 1.43 $\pm$ 0.21 &                 & 2.26 $\pm$ 0.88 \\
SW And    &   1878 & RRab & 0.442 &  9.6 & 1.77 $\pm$ 0.26 &                 & 1.48 $\pm$ 1.21 \\
XZ Cyg    &  96112 & RRab & 0.467 &  9.9 & 1.56 $\pm$ 0.23 & 1.67 $\pm$ 0.17 & 2.29 $\pm$ 0.84 \\
AV Peg    & 107935 & RRab & 0.390 & 10.4 & 1.53 $\pm$ 0.23 &                 & 2.28 $\pm$ 1.72 \\
V4424 Sgr &  97923 & RRab & 0.425 & 10.2 & 1.66 $\pm$ 0.25 &                 & 0.92 $\pm$ 1.94 \\
RX Eri    &  22442 & RRab & 0.587 &  9.7 & 1.83 $\pm$ 0.28 &                 & 1.50 $\pm$ 1.12 \\
BH Peg    & 112994 & RRab & 0.641 & 10.6 & 1.40 $\pm$ 0.22 &                 & 0.31 $\pm$ 1.82 \\
BN Vul    &  95702 & RRab & 0.594 & 10.7 & 1.45 $\pm$ 0.23 &                 & 6.09 $\pm$ 2.24 \\
SU Dra    &  56734 & RRab & 0.660 &  9.7 & 1.43 $\pm$ 0.28 & 1.42 $\pm$ 0.16 & 0.20 $\pm$ 1.13 \\
 \\
\multicolumn{8}{c}{Type-II Cepheids} \\
VY Pyx       & 43736 & BL Her & 1.239  & 7.0   & 3.85 $\pm$ 0.28 & 6.44 $\pm$ 0.23 & 5.01 $\pm$ 0.44 \\
KT Com       & 66179 & W Vir  & 4.070  & 8.0   & 4.16 $\pm$ 0.66 &                 & 5.50 $\pm$ 0.73  \\ 
$\kappa$ Pav & 93015 & W Vir  & 9.078  & (5.0) &                 & 5.57 $\pm$ 0.28 & 6.52 $\pm$ 0.77 \\ 

\hline

  \end{tabular}
  }
 \end{center}
\end{table}

There is only one classical Cepheid with an accurate parallax in GDR1, CK Cam.
Table~\ref{Tab-CEP} lists that star and the 12 stars which have an {\it HST} based parallax from
Benedict et al. (2007), Riess et al. (2014) and Casertano et al. (2016).
Most are too bright to be included in GDR1.
The two fainter stars suggest that the parallaxes derived using the new WFC3 scanning technique will be competitive beyond GDR2.

\begin{table}
  \begin{center}
  \caption{Data on classical Cepheids.}
  \label{Tab-CEP}
 {\scriptsize
  \begin{tabular}{|l|c|c|c|c|}\hline 

   Name      &  V  & $\pi \pm \sigma_{\pi}$ &   $\pi \pm \sigma_{\pi}$ &   $\pi \pm \sigma_{\pi}$ \\ 
             &     &    (mas, HST)         &    (mas, Hipparcos)     &   (mas, GDR1)          \\ \hline
$\beta$ Dor  & 3.5 & 3.14 $\pm$ 0.16 & $3.64 \pm 0.28$ &   \\
$\delta$ Cep & 3.7 & 3.66 $\pm$ 0.15 & $3.81 \pm 0.20$ &  \\ 
FF Aql       & 4.7 & 2.81 $\pm$ 0.18 & $2.05 \pm 0.34$ &  $1.64 \pm 0.89$ \\
$l$ Car      & 3.2 & 2.01 $\pm$ 0.20 & $2.06 \pm 0.27$ &  \\ 
RT Aur       & 5.3 & 2.40 $\pm$ 0.19 & $-0.23 \pm 1.01$ & \\ 
T Vul        & 5.5 & 1.90 $\pm$ 0.23 & $2.31 \pm 0.29$ &  \\ 
Y Sgr        & 5.1 & 2.13 $\pm$ 0.29 & $3.73 \pm 0.32$ &  \\ 
X Sgr        & 4.0 & 3.00 $\pm$ 0.18 & $3.39 \pm 0.21$ &  \\
$\zeta$ Gem  & 3.8 & 2.78 $\pm$ 0.18 & $2.71 \pm 0.17$ &  \\ 
W Sgr        & 4.3 & 2.28 $\pm$ 0.20 & $2.59 \pm 0.75$ &  \\ 

SS CMa       & 9.9 & 0.348 $\pm$ 0.038 &               &  $0.69 \pm 0.23$ \\
SY Aur       & 9.1 & 0.428 $\pm$ 0.054 &               &  $0.69 \pm 0.25$ \\ 

CK Cam       & 7.6 &                   &  $-0.59 \pm 1.13$ &  $1.56 \pm 0.25$ \\ \hline
  \end{tabular}
  }
 \end{center}
\end{table}

There is a very large number of Mira and SR variables listed in the GCVS, but since these stars are intrinsically bright 
only one has an accurate parallax, the anonymous SRb variable V375 And.
Whitelock \& Feast (2000) and Whitelock et al. (2008) studied Miras and Mira-like variables and derived the $K$ band $PL$ relation.
Table~\ref{Tab-MSR} lists 8 stars with relative parallax error $<0.16$ in {\it Hipparcos} data.
I also added R Dor, the star with the largest angular diameter on the sky (see column~6).
This is a relevant factor for these very large giants and supergiants, that have large convective cells.
Chiavassa et al. (2011) show that in a star like Betelgeuse the photocentre shifts by a noise characterised 
by a standard deviation of the order of 0.1 AU.
They find that in the worst situation, the degradation of the astrometric fit caused by 
this photocentric noise will be noticeable up to about 5 kpc for the brightest supergiants.

The effect could possibly be present in Cepheids as well but should be almost an order of magnitude smaller.
The largest Cepheid is $l$ Car with a mean angular diameter of $\sim3$ mas (Kervella et al. 2004) comparable to its parallax.
Others are smaller; see Table~12 in Groenewegen (2013) for predicted angular diameters and references to measured ones.


\begin{table}
  \begin{center}
  \caption{Data on Mira and SR variables.}
  \label{Tab-MSR}
 {\scriptsize
  \begin{tabular}{|l|c|c|c|c|c|c|}\hline 

   Name   & Type & V (GCVS)    &   $\pi \pm \sigma_{\pi}$ & $\pi \pm \sigma_{\pi}$ & $\theta$ & Reference for $\theta$ \\
          &      & (max - min) &      (mas, Hipparcos)   &      (mas, GDR1)      &  (mas)   & \\  \hline
V375 And  & SRb  & 7.0 - 7.2  &   2.35 $\pm$ 0.54        &  2.91 $\pm$ 0.46      &          &  \\ 
\hline
$o$ Cet   & M    & 2.0 - 10.1 & 10.91 $\pm$ 1.22 & & $33.6 \pm 3.5$ & Whitelock \& Feast (2000) \\ 
L$_2$ Pup & SRb  & 2.6 -  6.2 & 15.61 $\pm$ 0.99 & & $17.9 \pm 1.6$ & Kervella et al. (2014) \\ %
R Car     & M    & 3.9 - 10.5 &  6.34 $\pm$ 0.81 & & $\sim 20$      & Ireland et al. (2004) \\ %
R Leo     & M    & 4.4 - 11.3 &  9.01 $\pm$ 1.42 & & $37.4 \pm 2.3$ & Whitelock \& Feast (2000) \\ 
R Hya     & M    & 3.5 - 10.9 &  8.24 $\pm$ 0.92 & & $28.7 \pm 3.3$ & Whitelock \& Feast (2000) \\ 
W Hya     & SRa  & 7.7 - 11.6 &  9.59 $\pm$ 1.12 & & $45   \pm 4$   & Whitelock \& Feast (2000) \\ 
W Cyg     & SRb  & 6.8 -  8.9 &  5.72 $\pm$ 0.38 & & $11.5 \pm 0.4$ & Dyck et al. (1996)    \\ %
R Cas     & M    & 4.4 - 13.5 &  7.95 $\pm$ 1.03 & & $24.9 \pm 2.9$ & Whitelock \& Feast (2000) \\ %
\hline
R Dor     & SRb  & 4.8 -  6.6 & 16.02 $\pm$ 0.69 & & $57 \pm 5$    & Whitelock \& Feast (2000) \\

\hline

  \end{tabular}
  }
 \end{center}
\end{table}

\vspace{-2.0mm}
\section{GDR1}

Several papers have used GDR1 data in order to study the classical variables.
Two important ones have already been mentioned, (1) Gaia collaboration et al. (2017) that analysed
the parallax data in TGAS for known RRL, T2C, CEP and derived the zeropoint of various $PL$ relations 
(see Table~\ref{Tab-PL}),
and (2) Clementini et al. (2016) that analysed classical variables in the south ecliptic pole data.

Casertano et al. (2017) used the 212 Cepheids from van Leeuwen et al. (2007) with $VIJH$ data to
construct the $m_{\rm H} = m_{160} -0.3861 (m_{555} - m_{814})$ magnitude and compare the TGAS parallax
to the photometric parallax calculated from their adopted absolute calibration
$M_{\rm H} = -2.77 -3.26 \log P$.
They find that "the parallaxes are in remarkably good global agreement with the predictions, and there
is an indication that the published errors may be conservatively overestimated by about 20\%. 
Our analysis suggests that the parallaxes of 9 Cepheids brighter than G = 6 may be systematically underestimated".

Gould et al. (2016) use a similar approach and compare TGAS to photometrically determined parallaxes for
100 RRab stars using the $K$ band $PL$ relation, and find that the errors in TGAS are overestimated. 
The error in parallax quoted in GDR1 are inflated compared to the formal parallax uncertainty 
(Eq.~4 and Appendix~B in Lindegren et al. 2016), 
$\sigma_{\rm tgas} (\pi) = \sqrt{ (A \sigma_{\rm int})^2 + \sigma_0^2}$, where
$(A, \sigma_0)$ = (1.4, 0.2) is used in GDR1.
Gould et al. propose that (1.1, 0.12) is more appropriate.

\vspace{-2.0mm}
\section{Outlook}

The first data release of {\it Gaia} has shown the potential impact that this data can have on the calibration of
the distance scale, and that the community seems ready for GDR2!
The number of classical variables that can be expected is huge.  
From Table~20 in Robin et al. (2012) "Gaia Universe model snapshot" one can deduce that in the 
full catalog ($G<20$), or at the bright end ($G<12$), where additional abundance and detailed RV monitoring 
data will be available, one may expect 80~000 (400) RRab, 6500 (2200) classical Cepheids, and 40~000 (18~000) Mira variables.
Eyer \& Cuypers (2000) quote similar numbers.

%

In GDR2 one may already expect significant better precision in the parallaxes, as well as
time series of the $G$ band, and of the integrated BP and RP bands, providing colour information.
There already may be an all-sky release and characterisation of RRL with sufficient epochs.

As became clear from GDR1, an important issue is the bright limit, that is currently near $G= 6$ and that 
has a huge impact on the availability of parallax data for the best known classical Cepheids with accurate {\it HST} parallaxes.
Efforts are ongoing to bring this limit to $G= 3$ (Sahlmann et al. 2016), or even slightly brighter (Sahlmann et al., this volume).
An alternative route where {\it Gaia} could also contribute is to study Cepheids in clusters (Anderson et al. 2013; Chen et al. 2015).
The well known Cepheids $\delta$ Cep and $\zeta$ Gem are located in clusters (Majaess et al. 2012a,b) that can
provide alternative distances via main-sequence fitting.

\begin{figure}[b]
 \vspace*{-1.0 mm}
\begin{center}
 \includegraphics[width=98mm]{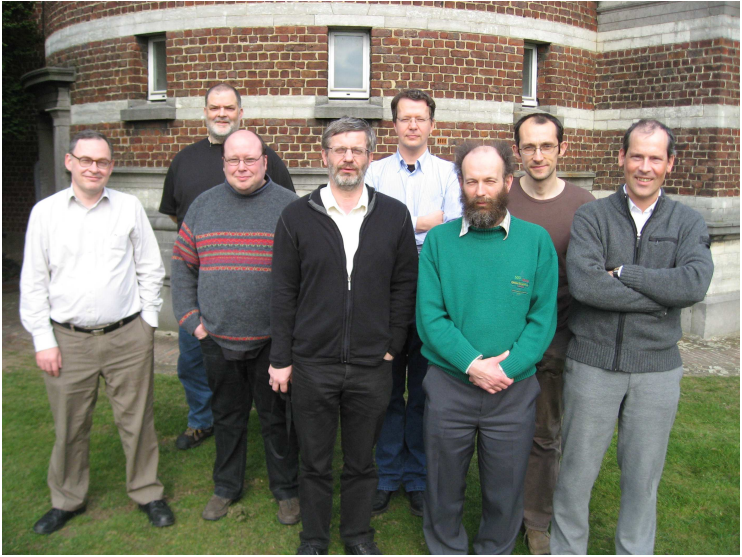} 
 \caption{This contribution is dedicated to the memory of Jan Cuypers (1956-2017) who died unexpectedly 
on the last day of February. Not only was he the head of the outreach department of the 
Royal Observatory of Belgium, and head of the Astronomy and Astrophysics department, 
Jan was heavily involved in {\it Gaia} in the context of DPAC Coordination Unit 7 on period determination 
and variable star classification.
The picture was taken in 2010. 
It shows Jan fourth from the left with his colleagues from the Royal Observatory involved in {\it Gaia}.}
   \label{fig1}
\end{center}
\end{figure}

\vspace{-2.0mm}
\begin{acknowledgements}
This work has made use of data from the European Space Agency (ESA)
mission {\it Gaia} (\url{www.cosmos.esa.int/gaia}), processed by
the {\it Gaia} Data Processing and Analysis Consortium (DPAC,
\url{www.cosmos.esa.int/web/gaia/dpac/consortium}). Funding
for the DPAC has been provided by national institutions, in particular
the institutions participating in the {\it Gaia} Multilateral Agreement.
\end{acknowledgements}





\end{document}